\documentclass[a4paper,11pt]{article}

\usepackage{textcomp}
\usepackage{amsfonts}
\usepackage[dvips]{graphicx}
\usepackage{amsmath}
\usepackage{amssymb}

\title{\textbf{Dynamical Systems on Three Manifolds\\
Part I: Knots, Links and Chaos}}

\author{Yi SONG, Stephen P. BANKS and David DIAZ\\
Department of Automatic Control and Systems Engineering,\\
University of Sheffield, Mappin Street,\\
Sheffield, S1 3JD.\\
e-mail: s.banks@sheffield.ac.uk}

\begin{document}

\maketitle

\newtheorem{theorem}{Theorem}[section]
\newtheorem{corollary}{Corollary}[section]
\newtheorem{definition}{Definition}[section]
\newtheorem{example}{Example}[section]

\begin{abstract}
In this paper, we give an explicit construction of dynamical systems
(defined within a solid torus) containing any knot (or link) and
arbitrarily knotted chaos. The first is achieved by expressing the
knots in terms of braids, defining a system containing the braids
and extending periodically to obtain a system naturally defined on a
torus and which contains the given knotted trajectories. To get
explicit differential equations for dynamical systems containing the
braids, we will use a certain function to define a tube neigbourhood
of the braid. The second one, generating chaotic systems, is
realized by modelling the \emph{Smale horseshoe}.
\end{abstract}

\section{Introduction}

Knot theory has been an important subject in its own right for a
long time (see [Kauffman, 1987]), and recently a great deal has been
written on the connections between knot theory and dynamical systems
[Ghrist et al, 1997]. The key idea is this: a closed (periodic)
orbit in a three-dimensional flow is an embedding of the circle,
$S^{1}$, into the three-manifold that constitutes the state space of
the system, hence it is a knot (see [Kauffman, 1987]). Hence,
periodic solutions of dynamical systems may be knotted or linked,
and, in fact, a chaotic system contains any knot and link ([Birman
and Williams, 1983]). A simple approach to obtaining a (non-chaotic)
system which contains an arbitrary knot (even a wild knot) is given
in [Banks and Diaz, 2004].

It is widely known that any knot can be expressed in terms of
braids, so in the first part of this paper we propose to show how to
write down general explicit differential equations for these braids
over a finite time interval, and then by making the vector field
periodic, we can glue the two ends of the phase space at successive
periodic time points together (see \textbf{Fig} \ref{braid
representation}), which will give us the desired knot embedded
within a solid torus.

\begin{figure}[!hbp]
\begin{center}
\includegraphics[width=2.5in]{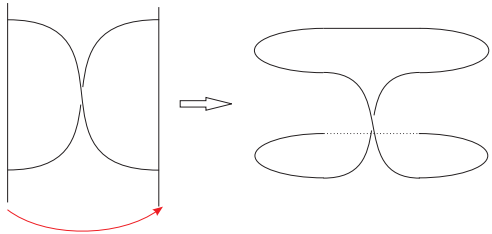}
\caption{Constructing knots from braids by gluing the two ends
together} \label{braid representation}
\end{center}
\end{figure}

\noindent This is achieved by using $C^{\infty }$ functions to make
the twists in the braids. We shall find that any braid in a solid
torus is given by an equation of the form
\[
\left\{
\begin{array}{l}
\dot{x}=\sum_{i}f_{i}(x,y,z) \\
\dot{y}=\sum_{i}g_{i}(x,y,z) \\
\dot{z}=c
\end{array}
\right.
\]
where $c$ is a constant and $f$ and $g$ are $C^{\infty }$ functions
(i.e. an infinitely differentiable functions) which are identically
1 on an interval $(-\infty ,a)$ and identically 0 on an interval
$(b,\infty )$, with $a<b$. From the equations above, our dynamical
system is defined by the sum of a sequence of $C^{\infty }$
functions (combined in a proper way). However, different strands of
the braids will certainly affect each other, which makes it
difficult to control the global behaviour of the whole system,
especially at the two ends, where we want the vector field to fit
together smoothly. This again involves the introduction of a
$C^{\infty }$ function to control the whole dynamical system so that
the vector fields corresponding to different strands in our braid
will not interact.

We shall introduce the algorithm to express knots in terms of braids
in the next section. This result is certainly well known, but we
include this for the convenience of the reader and to fix the
notation and ideas for the rest of the work. Then in \S 3 we outline
the main idea of $C^{\infty }$ functions that we need to generate
dynamical systems for braids, which will glue together and be knots,
as will be stated in \S 4. Finally, we will consider how to use
similar ideas to create chaotic systems (containing no homoclinic
orbits) in \S 5.

In the second part of the paper we shall consider a more general
approach to the study of dynamical systems on three manifolds, using
\emph{Heegaard splittings} and the theory developed recently in
[Banks \& Song, 2006] for the structure of general dynamical systems
on surfaces, using \emph{automorphic function} theory.

\section{Relation between Knots and Braids}

We shall generate systems containing any given knot by expressing
the knot in the form of a closed braid. Then we determine a system
with a periodic vector field which contains the desired braid. First
we outline \emph{Alexander's theorem} relating braids to knots.

\begin{figure}[!hbp]
\begin{center}
\includegraphics[width=4in]{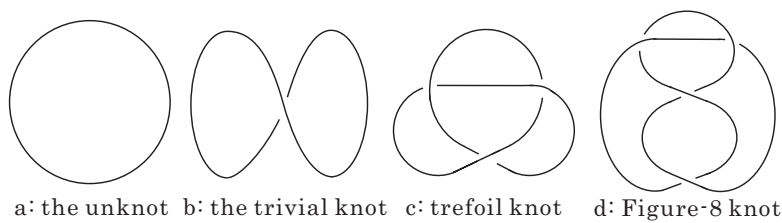}
\caption{The simplest knots} \label{simple knot}
\end{center}
\end{figure}

A \emph{knot} is a smooth embedding of the circle $S^{1}$ in
$\mathbb{R}^{3}$ (see \textbf{Fig} \ref{simple knot}), and a
\emph{link} is a finite disjoint collection of knots. In this paper
we only consider knots, but the ideas generalize easily to links.

\begin{figure}[!h]
\begin{center}
\includegraphics[width=2.5in]{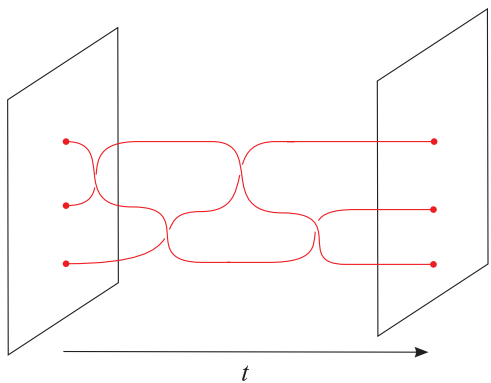}
\caption{A braid in 3-space} \label{braid in 3-space}
\end{center}
\end{figure}

Knot theory is studied conveniently in terms of braids, which were
originally introduced by [Artin, 1925]. By definition, an
\emph{m-strand braid} is a set of $m$ non-intersecting smooth paths
connecting $m$ points on one horizontal line to $m$ points on
another horizontal line (below the first one) - see [Kauffman,1987]
and \textbf{Fig} \ref{braid in 3-space}. If we glue the
corresponding left and right hand side of the braid together
respectively, we get the so-called \emph{closure} of a braid. From
\textbf{Fig} \ref{braid in 3-space}, we see that it is actually a
knot. More generally, the closure of a braid is a link. Usually
closures of braids are taken to be oriented, all strands of the
braid are oriented from left to right in this paper.

\begin{theorem}
(Alexander's theorem) Each link can be represented as
the closure of a braid.  $\Box $
\end{theorem}

Since a link is a smooth embedding of several disjoint circles in
$\mathbb{R} ^{3}$, it is actually composed of several knots.
Consequently, we have the following corollary.

\begin{corollary}
Each knot can be represented as the closure of a braid. $\Box $
\end{corollary}

Now we give an algorithm to construct braid from a given knot. We
will illustrate the method by the use of a \emph{figure-8 knot} --
the general case will then be clear. (see \textbf{Fig} \ref{figure8
knot representation}).

\begin{figure}[!hbp]
\begin{center}
\includegraphics[width=3in]{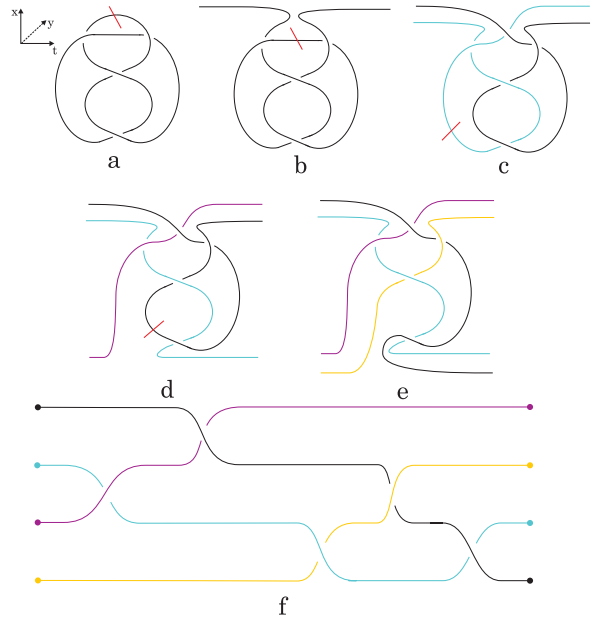}
\caption{Construction of a braid from the \emph{figure-8} knot}
\label{figure8 knot representation}
\end{center}
\end{figure}

Since a knot is essentially a closed loop, while the braid
construction from a knot has at lease one strand (in the
\emph{unknot} case), and each strand has two ends, basically what we
should do is to cut the knot in a proper way so that it will turn
into several strands which give us the correct braid representation.

First, we choose one part of the given knot which is away from the
crossings, cut the diagram at a point from this part and straighten
the two ends resulted from the cut, which gives us the first strand
of the braid. Obviously it is better to cut the top or the bottom of
the knot, as shown in \textbf{Fig} \ref{figure8 knot representation}
(a) and (b).

As we know, the only knot that has just a one-string braid is the
\emph{unknot}, which implies that all the other knots have more than
one strand in the corresponding braid construction. This means in
order to build up the braid representation, there is a need to cut
the diagram again to get the remaining strands. Self-crossings of
any strand of the braid can be removed by \emph{Reidemeister moves}.
So the next step will be choosing a part of the resulting diagram
which is before or after a self-crossing of the one-strand from the
first step, cut it and straighten the diagram as we did previously.
A little care must be taken at this place, since a knot is just one
circle and so for each strand in the braid representation, we must
guarantee that it starts at one end and finishes at the other. The
solution to this is simple: we just need to add another crossing if
necessary to ensure that a strand comes in from one side must go out
from the other side. (see \textbf{Fig} \ref{figure8 knot
representation} (c) and (d) for an illustration) Meanwhile, for each
strand, the starting and ending points cannot share the same
$x$-value or $y$-value (according to the coordinate as shown in
\textbf{Fig} \ref{figure8 knot representation}), otherwise the braid
represents a multi-component link instead of a knot.

Perform this operation repeatedly until all the strands in the braid
have no self-crossings, as shown in \textbf{Fig} \ref{figure8 knot
representation} (e).

Finally, rearrange the braid so that there is at most one twist at
each vertical strip (as shown in \textbf{Fig} \ref{figure8 knot
representation} (f)).

Eventually we get a braid representation of the knot. Of course,
there are an infinite number of braid representations of a given
knot; however, we will choose the simplest one to study in this
paper.

\section{$C^{\infty }$ Functions}

Now we shall give a brief resume of the theory of $C^{\infty }$
functions which we need in the next section. All the results are
well known, and can be found, for example in [Helgason, 1978].

Let $\mathbb{R}^{m}$ and $\mathbb{R}^{n}$ denote two
\emph{Euclidean} spaces of $m$ and $n$ dimensions, respectively. Let
$S$ and $S'$ be open subsets of $\mathbb{R}^{m}$ and
$\mathbb{R}^{n}$, respectively, and suppose $\psi $ is a mapping
from $S$ to $S'$.

\begin{definition}
\emph{The mapping $\psi $ is called \emph{differentiable} if the
coordinates $y_{j}(\psi (p))$ of $\psi (p)$ are differentiable
functions of the coordinates $x_{i}(p)$, $p\in S$.}
\end{definition}

\begin{definition}
\emph{The mapping $\psi $ is called \emph{analytic} if for each
point $p\in S$ there exists a neighbourhood $U$ of $p$ and $n$ power
series $P_{j}$ $(1\leq j\leq n)$ in $m$ variables such that
$y_{j}(\psi (q))=P_{j}(x_{1}(q)-x_{1}(p),\cdots ,x_{m}(q)-x_{m}(p))$
$(1\leq j\leq n)$ for $q\in U$.}
\end{definition}

\begin{definition}
\emph{A differentiable mapping $\psi :O\rightarrow O'$ is called a
\emph{diffeomorphism} of $O$ and $O'$ if, $\psi$ is one-to-one and
onto, and the inverse mapping $\psi ^{-1}$ is differentiable.}
\end{definition}

For an analytic function on $\mathbb{R}^{m}$, if it vanishes on an
open set, then it is identically zero. However, for general
differentiable functions and in particular $C^{\infty }$ functions,
the situation is completely different.

\begin{definition}
\emph{If $A$ and $B$ are two disjoint subsets of $\mathbb{R}^{m}$,
then there exists an infinitely differentiable function $\varphi$
which is identically 1 on $A$ and identically 0 on $B$. To emphasize
the dependence on $A$ and $B$ we often write this as $\varphi
(x;A,B)$.}
\end{definition}

Obviously such a function is non-analytic, since it is identically 0 or 1
for a continuous interval; but it is infinitely differentiable, which makes
it very useful in the next section.

\begin{figure}[!hbp]
\begin{center}
\includegraphics[width=2in]{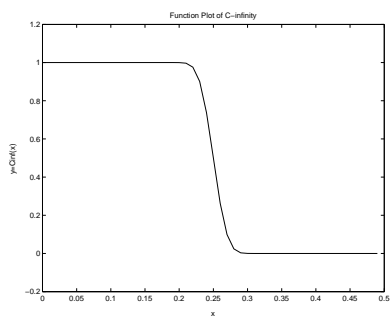}
\caption{Function plot of $C^{\infty}$} \label{c infinity function}
\end{center}
\end{figure}

The standard procedure for constructing such a $C^{\infty }$
function is as follows: Let $0<a<b$ and consider the function $f$ on
$\mathbb{R}$ defined by

\begin{equation}
f(x)=\left\{
\begin{array}{ll}
\exp ( \frac{1}{x-b}-\frac{1}{x-a}) & \text{if }a<x<b \\
0 & \text{otherwise.}
\end{array}
\right.
\end{equation}
Then $f$ is differentiable and the same holds for the function
\begin{equation}
F(x)=\frac{\int_{x}^{b}f(t)dt}{\int_{a}^{b}f(t)dt},
\end{equation}
which has value 1 for $x\leq a$ and 0 for $x\geq b$. The $C^{\infty
}$ function $\varphi $ defined on $\mathbb{R}^{m}$ is
\begin{equation}
\varphi (x_{1},\cdots ,x_{m})=F(x_{1}^{2}+\cdots +x_{m}^{2}).
\end{equation}
It can be seen that $\varphi $ is differentiable and has values 1
for $ x_{1}^{2}+\cdots +x_{m}^{2}\leq a$ and 0 for $x_{1}^{2}+\cdots
+x_{m}^{2}\geq b$, by a slight abuse of notation we shall write it
as $ \varphi (x;a,b)$. (see \textbf{Fig} \ref{c infinity function})

In fact we can approximate it by just using an exponential function,
say $y=\exp (-x^{20})$. However, for exact matching at the
boundaries we require a function which is constant on certain
regions of space.

\section{Dynamical Systems for Braids}

We now consider dynamical systems which contain any braids
constructed from some given knots. Using the coordinate system shown
in \textbf{Fig} \ref{figure8 knot representation}, we notice that a
braid is composed of several strands and twists, with at most one
twist at a certain vertical strip (an interval of the $t$
coordinate). Hence if we can find a dynamical system which gives us
the twist, then it just remains to repeat the process to give the
appropriate number of twists for the complete braid.

Each strand in the braid is defined by a set of equations of the form

\begin{eqnarray*}
\dot{x} &=& G_{1}(x,y,t) \\
\dot{y} &=& G_{2}(x,y,t) \\
\dot{z} &=& G_{3}(x,y,t)
\end{eqnarray*}

\noindent where $G_{1},G_{2}$ and $G_{3}$ are some functions of $x$,
$y$ and $t$. Normally we take $G_{3}$ to be a constant so the
$z$-axis is effectively the time axis and periodicity of $G$ with
respect to $t$ will lead to a system with the required knot.

\begin{figure}[!hbp]
\begin{center}
\includegraphics[width=4in]{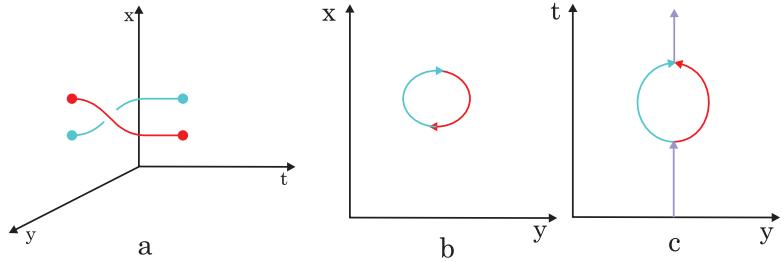}
\caption{One twist of a braid} \label{one twist}
\end{center}
\end{figure}

\textbf{Fig} \ref{one twist} shows a twist projected onto three
different planes, namely the $xz$-, $xy$-, and $zy$-planes,
respectively.Because of the need to glue the braid together to get a
knot, we assume that at the two ends, all the strands are parallel
to the $z$-axis, (mathematically speaking, we need
$\dot{x}=0$,$\dot{y}=0$) which makes the vector fields at these
connecting points match at the periodic points. We now give explicit
equations for these strings.

As shown in \textbf{Fig} \ref{one twist}, in the $xz$-plane
projection, the shape of the red strand is that of the $C^{\infty }$
function. After studying the change of the vector field, $\dot{x}$
,with respect to $t$, we get
\begin{equation} \label{first cinfinity equation}
\dot{x}=\varphi (t;a,b)-\varphi (t;b,c),\qquad(a<b<c).
\end{equation}
In case of an ascending strand instead of the descending one, we
have
\begin{equation} \label{second cinfinity equation}
\dot{x}=\varphi (t;b,c)-\varphi (t;a,b),\qquad(a<b<c).
\end{equation}
In the $xy$-plane, the transformation group acts as a circle, which
brings the top one to the bottom and bottom to the top without
intersection. In the $zy$-plane, the trajectory is a semicircle in
the middle plus two straight lines at the two ends. A proper
combination of $C^{\infty }$ functions will give any desired link.
Thus, for an over-crossing, such as the red one in \textbf{Fig}
\ref{one twist}, we have
\begin{equation}
\dot{y}=\varphi (t;b,c)-\varphi (t;a,b)+\varphi (t;c,d)-\varphi
(t;d,e)\qquad(a<b<c<d<e)
\end{equation}
while for an under-crossing, such as the blue one, it becomes
\begin{equation}
\dot{y}=\varphi (t;a,b)-\varphi (t;b,c)+\varphi (t;d,e)-\varphi
(t;c,d)\qquad(a<b<c<d<e)
\end{equation}
We assume that $\dot{z}$ is a constant and set $z=t$; then the
equation for one twist in a braid is

\begin{eqnarray}
\dot{x} &=&\left\{
\begin{array}{l}
\varphi (t;a,b)-\varphi (t;b,c) \\
\varphi (t;b,c)-\varphi (t;a,b)
\end{array}
\right. \qquad(a<b<c)  \nonumber \\
\dot{y} &=&\left\{
\begin{array}{l}
\varphi (t;b,c)-\varphi (t;a,b)+\varphi (t;c,d)-\varphi (t;d,e) \\
\varphi (t;a,b)-\varphi (t;b,c)+\varphi (t;d,e)-\varphi (t;c,d)
\end{array}
\right. \\
& & \qquad \qquad \qquad \qquad \qquad \qquad (a<b<c<d<e) \nonumber \\
\dot{z} &=&\text{constant}  \nonumber
\end{eqnarray}
where the choice is made depending on whether the twisted strand is
ascending or descending, under-crossing or over-crossing.

Then we can get the dynamical system for just one strand, it is of
the form

\begin{eqnarray}
\dot{x} &=& \sum_{i=1}^{p}\pm \left( \varphi (t;a_{i},b_{i})-\varphi
(t;b_{i},c_{i})\right) \;(a_{i}<b_{i}<c_{i})  \nonumber \\
\dot{y} &=&\sum_{i=1}^{p}\pm \left( \varphi (t;a_{i},b_{i})-\varphi
(t;b_{i},c_{i})+\varphi (t;d_{i},e_{i})-\varphi (t;c_{i},d_{i})\right)
\\
& & \qquad \qquad \qquad \qquad \qquad (a_{i}<b_{i}<c_{i}<d_{i}<e_{i}) \nonumber \\
\dot{z} &=&\text{constant}  \nonumber
\end{eqnarray}
where $p$ is total number of twists in this strand, the $\pm $ sign is taken
depending one whether at the corresponding twist $i$, the strand ascends or
descends, under-crosses or over-crosses.

\begin{figure}[!hbp]
\begin{center}
\includegraphics[width=3in]{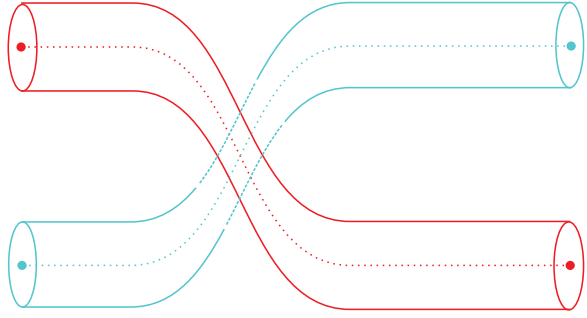}
\caption{Creating a tube around each existing strand} \label{tube}
\end{center}
\end{figure}

The next step will be to combine all the equations for different
strands together in a proper way so that we get a final one for the
whole braid. As shown below in \textbf{Fig} \ref{tube}, we create a
tube around each string, such that within this tube, all the
trajectories follow the middle strand, while outside it, the
dynamics are all zero, which means both $\dot{x}$ and $\dot{y}$ are
0.

By definition, an \emph{m-strand braid} is a set of $m$
non-intersecting smooth paths, so the key idea is this: as long as
the radius is small enough, there exists a tube around each string
that has no intersection with others. In this way, we effectively
get one equation for the whole braid while avoiding the interaction
between the stands.

This is also achieved by using the $C^{\infty }$ function of the form
\begin{equation} \label{cinfinity tube}
\varphi =\varphi ((x-x_{1})^{2}+(y-y_{1})^{2});a,b)
\end{equation}
where $(x_{1},y_{1})$ is the coordinate of the middle strand with respect to
different $t$ value, and $a$, $b$ need to be chosen carefully so that they
are small enough to avoid intersection with other tubes.

Consequently, the dynamical system of the braid is
\begin{eqnarray}
\dot{x} &=&\sum_{j=1}^{q}\varphi _{j}\cdot \dot{x}_{j}  \nonumber \\
\dot{y} &=&\sum_{j=1}^{q}\varphi _{j}\cdot \dot{y}_{j} \\
\dot{z} &=&\text{constant}  \nonumber
\end{eqnarray}
where $q$ is the total number of strands in the braid, $\varphi
_{j}$ is the tube function for the $j$th strand, and
$\dot{x}_{j},\dot{y}_{j}$ are the dynamics for the $j$th strand
obtained from \textbf{Equation} (\ref{cinfinity tube}). As before we
choose $\dot{z}$ to be a constant, so $z$ is like a time-axis.

\begin{example}
\emph{Consider the \emph{trefoil} knot shown in \textbf{Fig}
\ref{simple knot}; we shall give two braid presentations and the
corresponding dynamical systems for it.}

\emph{i) As shown in \textbf{Fig} \ref{trefoil 1}, the
\emph{trefoil} knot can be represented by a 2-strand braid.}
\emph{Hence the dynamics for strand $1$ is of the form}
\begin{figure}[!hbp]
\begin{center}
\includegraphics[width=3in]{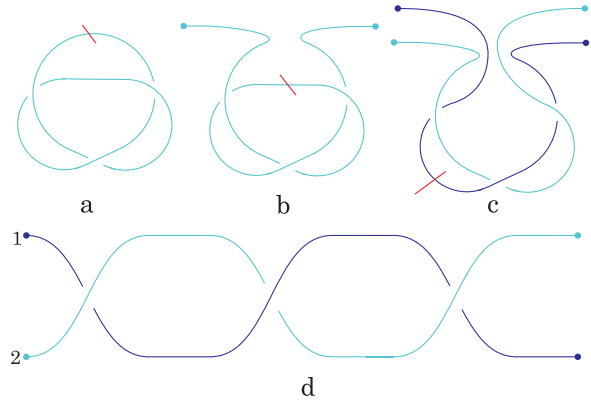}
\caption{Braid construction of the \emph{trefoil} knot -- method 1}
\label{trefoil 1}
\end{center}
\end{figure}
{\setlength\arraycolsep{1pt}
\begin{eqnarray}
\dot{x} &=&\varphi (t;a,b)-\varphi (t;b,c)+\varphi (t;e,f)-\varphi
(t;d,e)+\varphi (t;g,h)-\varphi (t;h,i)  \nonumber \\
\dot{y} &=&\varphi (t;a,\frac{a+b}{2})-\varphi (t;\frac{a+b}{2},b)+\varphi
(t;\frac{b+c}{2},c)-\varphi (t;b,\frac{b+c}{2})  \nonumber \\
&&+\varphi (t;\frac{d+e}{2},e)-\varphi (t;d,\frac{d+e}{2})+\varphi (t;\frac{%
e+f}{2},f)-\varphi (t;e,\frac{e+f}{2}) \\
&&+\varphi (t;g,\frac{g+h}{2})-\varphi (t;\frac{g+h}{2},h)+\varphi (t;\frac{%
h+i}{2},i)-\varphi (t;h,\frac{h+i}{2})  \nonumber \\
\dot{z} &=&\text{\emph{constant}}  \nonumber
\end{eqnarray}}
\emph{for some numbers $a,b,c,d,e,f,g,h,i$, while for strand $2$,
the equations are much the same except the change of plus/minus
sign. So let $(x_{1},y_{1})$ and $(x_{2},y_{2})$ stand for the $x$-
and $y$-value for strand $1$ and $2$ respectively, we can build up a
tube around each string, and finally get the dynamical system for
the braid, which is}

\begin{figure}[!hbp]
\begin{center}
\includegraphics[width=2.2in]{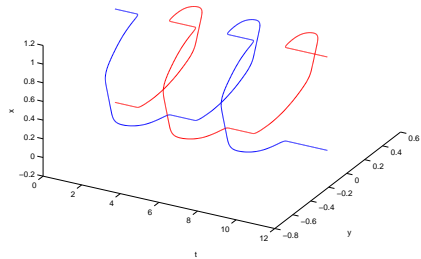}
\caption{A 2-strand braid generated from Matlab} \label{matlab
trefoil 1}
\end{center}
\end{figure}

\begin{eqnarray*}
\dot{x} &=&\varphi \left( (x-x_{1})^{2}+(y-y_{1})^{2};\xi _{1},\xi
_{2}\right) \times \\
&&\left( \varphi (t;a,b)-\varphi (t;b,c)+\varphi (t;e,f)-\varphi
(t;d,e)+\varphi (t;g,h)-\varphi (t;h,i)\right) \\
&&+\varphi \left( (x-x_{2})^{2}+(y-y_{2})^{2};\xi _{1},\xi _{2}\right) \times
\\
&&\left( \varphi (t;b,c)-\varphi (t;a,b)+\varphi (t;d,e)-\varphi
(t;e,f)+\varphi (t;h,i)-\varphi (t;g,h)\right) \\
\dot{y} &=&\varphi \left( (x-x_{1})^{2}+(y-y_{1})^{2};\xi _{1},\xi
_{2}\right) \times \\
&&\left( \varphi (t;a,\frac{a+b}{2})-\varphi (t;\frac{a+b}{2},b)+\varphi (t;%
\frac{b+c}{2},c)-\varphi (t;b,\frac{b+c}{2})\right. \\
&&+\varphi (t;\frac{d+e}{2},e)-\varphi (t;d,\frac{d+e}{2})+\varphi (t;\frac{%
e+f}{2},f)-\varphi (t;e,\frac{e+f}{2}) \\
&&+\left. \varphi (t;g,\frac{g+h}{2})-\varphi (t;\frac{g+h}{2},h)+\varphi (t;%
\frac{h+i}{2},i)-\varphi (t;h,\frac{h+i}{2})\right) \\
&&+\varphi \left( (x-x_{2})^{2}+(y-y_{2})^{2};\xi _{1},\xi _{2}\right) \times
\\
&&\left( \varphi (t;\frac{a+b}{2},b)-\varphi (t;a,\frac{a+b}{2})+\varphi
(t;b,\frac{b+c}{2})-\varphi (t;\frac{b+c}{2},c)\right. \\
&&+\varphi (t;d,\frac{d+e}{2})-\varphi (t;\frac{d+e}{2},e)+\varphi (t;e,%
\frac{e+f}{2})-\varphi (t;\frac{e+f}{2},f) \\
&&+\left. \varphi (t;\frac{g+h}{2},h)-\varphi (t;g,\frac{g+h}{2})+\varphi
(t;h,\frac{h+i}{2})-\varphi (t;\frac{h+i}{2},i)\right) \\
\dot{z} &=&\text{\emph{constant}}
\end{eqnarray*}
\emph{where $\xi _{1},\xi _{2}$ have to be chosen carefully to avoid
intersection of different tubes.}

\emph{Using Matlab, we get a plot of this 2-strand braid, as shown
in \textbf{Fig} \ref{matlab trefoil 1}.}

\emph{ii) By adding another cut as shown in \textbf{Fig}
\ref{trefoil 1} (c), we have a new braid presentation for the same
\emph{trefoil} knot, as in \textbf{Fig} \ref{trefoil 2}.}
\begin{figure}[!h]
\begin{center}
\includegraphics[width=3in]{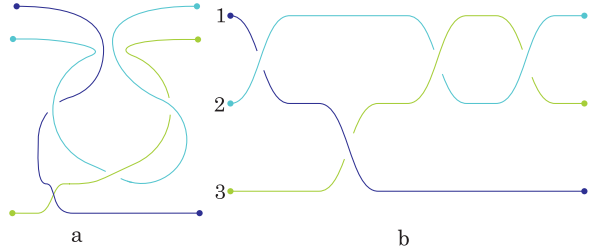}
\caption{Braid construction of the \emph{trefoil} knot -- method 2}
\label{trefoil 2}
\end{center}
\end{figure}

\emph{In the same manner as in method $1$, let
$(\dot{x}_{1},\dot{y}_{1},\dot{z}_{1}),(\dot{x}_{2},\dot{y}_{2},\dot{z}_{2}),(\dot{x}_{3},\dot{y}_{3},\dot{z}_{3})$
stand for the dynamics, $(x_{1},y_{1})$ ,$(x_{2},y_{2})$ and
$(x_{3},y_{3})$ for the $x$- and $y$-value of the three strings,
respectively. Then the dynamical system for this 3-strand braid is}

\begin{eqnarray}
\dot{x} &=&\varphi \left( (x-x_{1})^{2}+(y-y_{1})^{2};\xi _{1},\xi
_{2}\right) \times \dot{x}_{1}  \nonumber \\
&&+\varphi \left( (x-x_{2})^{2}+(y-y_{2})^{2};\xi _{1},\xi _{2}\right)
\times \dot{x}_{2}  \nonumber \\
&&+\varphi \left( (x-x_{3})^{2}+(y-y_{3})^{2};\xi _{1},\xi _{2}\right)
\times \dot{x}_{3}  \nonumber \\
\dot{y} &=&\varphi \left( (x-x_{1})^{2}+(y-y_{1})^{2};\xi _{1},\xi
_{2}\right) \times \dot{y}_{1} \\
&&+\varphi \left( (x-x_{2})^{2}+(y-y_{2})^{2};\xi _{1},\xi _{2}\right)
\times \dot{y}_{2}  \nonumber \\
&&+\varphi \left( (x-x_{3})^{2}+(y-y_{3})^{2};\xi _{1},\xi _{2}\right)
\times \dot{y}_{3}  \nonumber \\
\dot{z} &=&\emph{\text{constant}}  \nonumber
\end{eqnarray}
\emph{A picture for this 3-strand braid generated from Matlab is
shown in \textbf{Fig} \ref{trefoil 3}.}

\begin{figure}[!h]
\begin{center}
\begin{tabular}{cc}
\includegraphics[width=2.22in,height=2in]{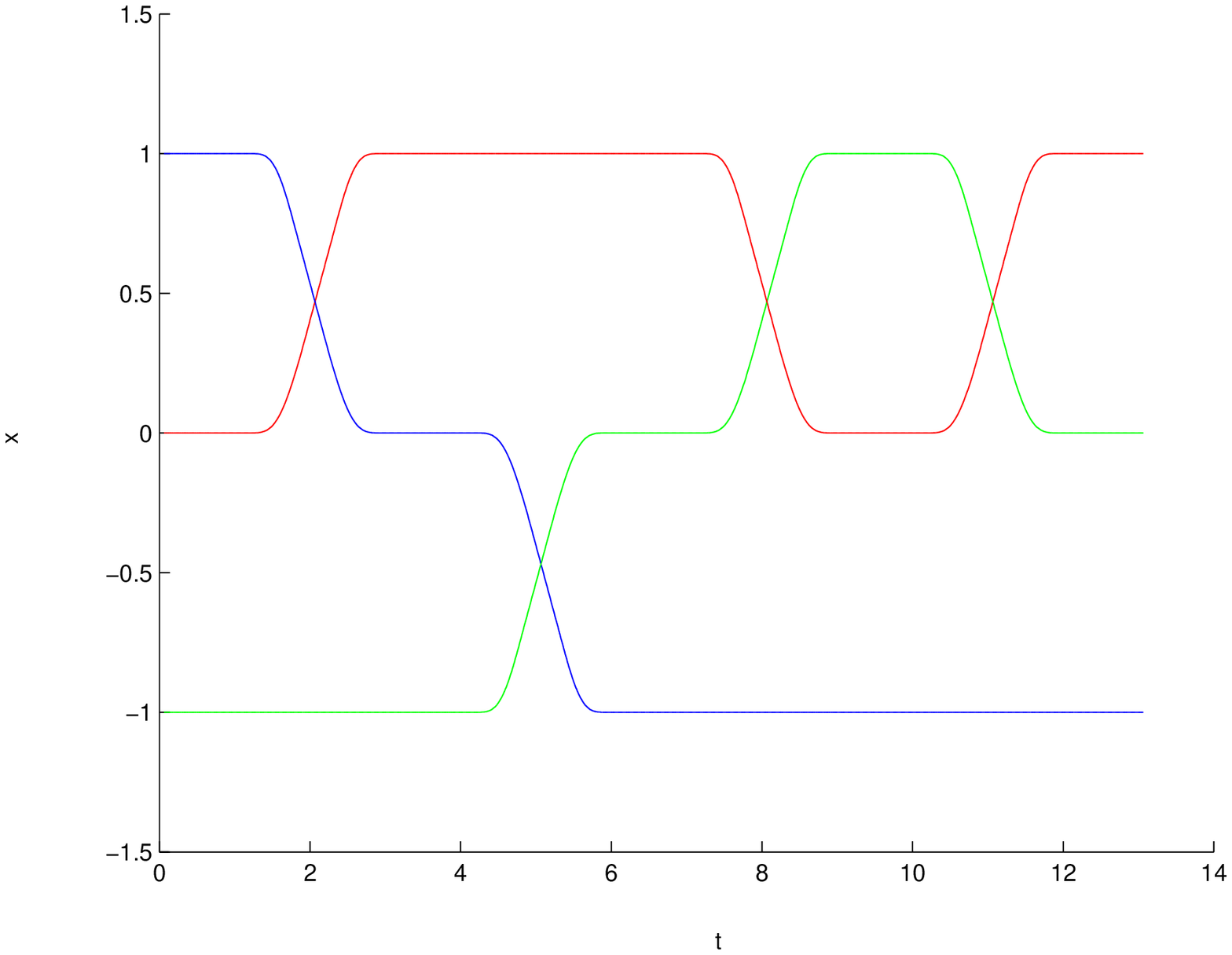} &
\includegraphics[width=2.22in,height=2.22in]{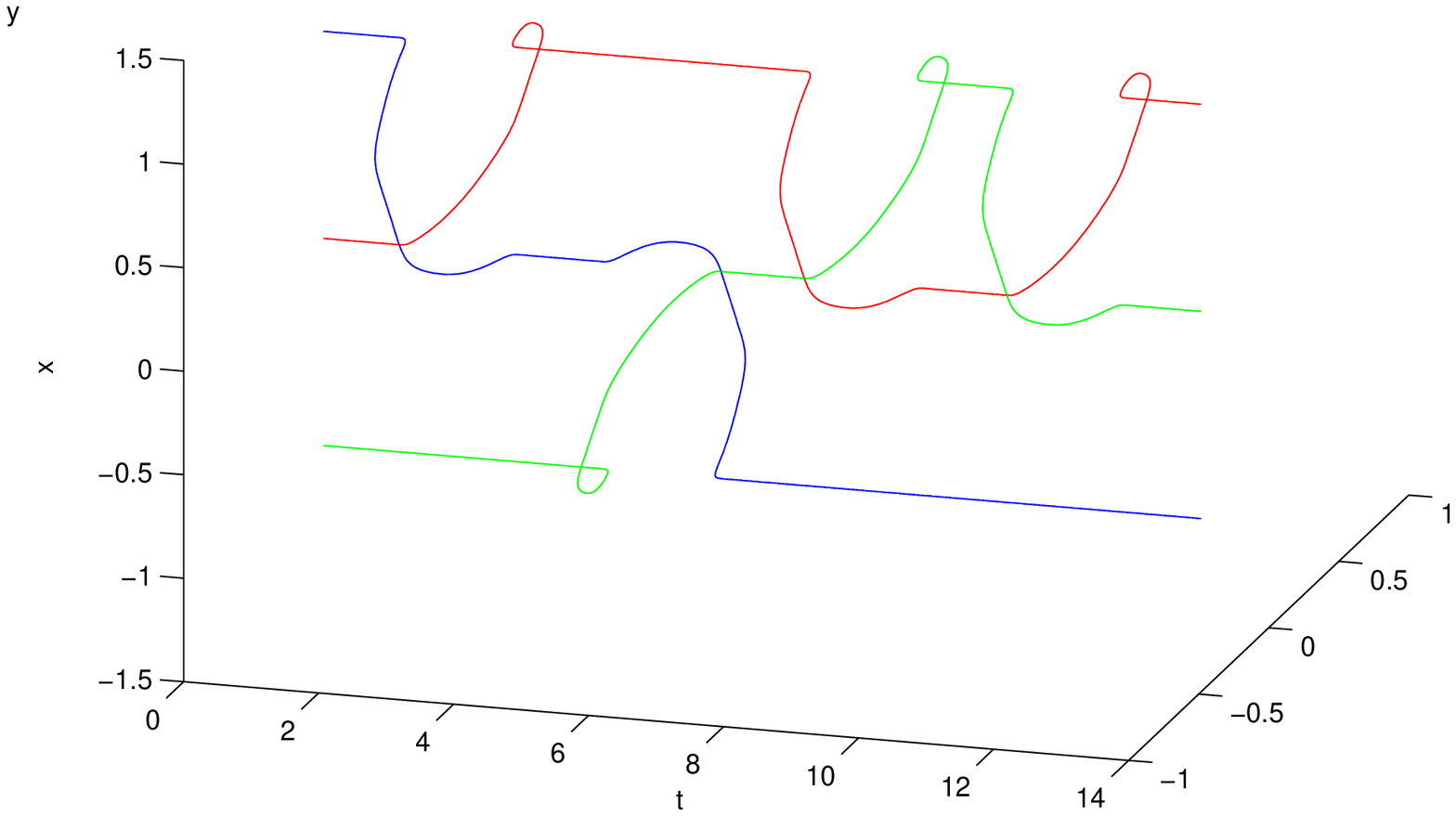}
\end{tabular}
\caption{A three strand braid presentation for the \emph{trefoil}
knot} \label{trefoil 3}
\end{center}
\end{figure}

\emph{After gluing the corresponding ends of the braid together, we
get the required knot situated in a solid torus - this is equivalent
to making the vector fields in the systems above periodic.}
\end{example}

\section{Chaotic Systems}

In this section we shall demonstrate how to obtain dynamical systems
with arbitrarily knotted chaos. We shall do this by making an
extension of the methods in the previous sections. In particular, we
shall need some elementary ideas from transformation group theory.
Thus, if $X$ is a topological space, and $ G $ is a group, we say
that $G$ is a \textbf{transformation group} on $X$ if there is a
continuous map $\varphi :G\times X\rightarrow X$ such that

(i) $\varphi (g,\varphi (h,x))=\varphi (gh,x)$ for all $g,h\in G,$ and all $%
x\in X$

(ii) $\varphi (e,x)=x$ for all $x\in X$, where $e$ is the
\emph{identity} of $G$.

\noindent We usually write $gx$ for $\varphi (g,x)$. If $G$ is a
subset of $Gl(n)$ (the general linear group), we call $G$ a
\textbf{linear transformation group}.

Consider now a process for modifying a given dynamical system
\[
\dot{x}=f(x,t)
\]
by a given function $t\rightarrow G(t)$ where $G(t)$ is an element of a
(linear) transformation group for each $t$. We define
\[
y(t)=G(t)x(t).
\]
Then
\begin{eqnarray}
\dot{y} &=&\dot{G}x+G\dot{x}  \nonumber \\
&=&\dot{G}G^{-1}y+Gf(x,t) \\
&=&\dot{G}G^{-1}y+Gf(G^{-1}y,t) \stackrel{\bigtriangleup}{=}
\tilde{G}(y,t) \nonumber
\end{eqnarray}

\begin{theorem} \label{simple}
Suppose that the vector field $(x,t)\rightarrow
f(x,t)$ (defined on a subset $U\subseteq \mathbb{R}^{n}$) is
periodic in $t$, with period $\pi$, and that the map $t\rightarrow
G(t)$, where $G(t)$ belongs to some linear transformation group on
$U$, is such that the vector field
\[
(y,t)\rightarrow \dot{G}(t)G^{-1}(t)y+G(t)f(G^{-1}(t)y,t)
\]
is also periodic in $t$ with period $\pi$, then the system
\[
\left\{
\begin{array}{l}
\dot{y}=\tilde{G}(y,t) \\
\dot{z}=2\sqrt{1-z^{2}}
\end{array}
\right. , \qquad y\in U,\;z(0)=0
\]
is naturally defined on the torus $U\times [0,1]/\sim $ where $\sim $ is the
equivalence relation
\[
(u,t)\sim (v,t)
\]
if and only if $u=v$ and $t=0$ or $\pi$.
\end{theorem}

\noindent \textbf{Proof.} The proof follows from the above
discussion and the fact that the unique solution of the equation
\[
\dot{z}=2\sqrt{1-z^{2}},\qquad z(0)=0
\]
is
\[
z(t)=\sin 2t
\]
which is periodic with period $\pi$.  \qquad $\Box$

\begin{example}
\emph{Consider the trivial system}
\[
\begin{array}{l}
\dot{x}_{1}=0 \\
\dot{x}_{2}=0
\end{array}
,\qquad 0\leq t\leq \pi
\]
\emph{defined on the disk $\{\left\| x\right\| <1\}$, and let $G$ be
the orthogonal group $O(2,\mathbb{R})$. Then if}
\[
G(t)=\left(
\begin{array}{ll}
\cos t & \sin t \\
-\sin t & \cos t
\end{array}
\right)
\]
\emph{(i.e. counterclockwise rotation through $t$), we have}
\begin{eqnarray*}
\dot{y}(t) &=&\dot{G}(t)G^{-1}(t)y(t)+G(t)\cdot \left(
\begin{array}{l}
0 \\
0
\end{array}
\right) \\
&=&\left(
\begin{array}{ll}
-\sin t & \cos t \\
-\cos t & -\sin t
\end{array}
\right) \cdot \left(
\begin{array}{ll}
\cos t & -\sin t \\
\sin t & \cos t
\end{array}
\right) \cdot y(t) \\
&=&\left(
\begin{array}{ll}
0 & 1 \\
-1 & 0
\end{array}
\right) \cdot y(t).
\end{eqnarray*}
\emph{Hence the system}
\begin{eqnarray*}
\dot{y}(t) &=&\left(
\begin{array}{ll}
0 & 1 \\
-1 & 0
\end{array}
\right) \cdot y(t) \\
\dot{z} &=&2\sqrt{1-z^{2}},\qquad z(0)=0
\end{eqnarray*}
\emph{has trefoil knot solutions (see \textbf{Fig} \ref{cylinder}).}

\begin{figure}[!hbp]
\begin{center}
\includegraphics[width=3in]{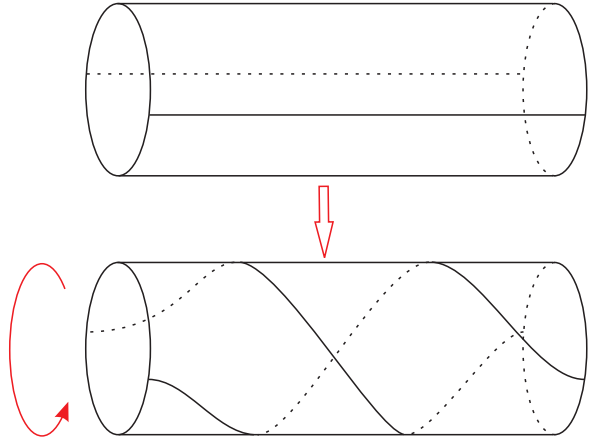}
\caption{Forming a \emph{trefoil} knot} \label{cylinder}
\end{center}
\end{figure}

\end{example}

This is, of course, a trivial example and to be useful we often
require to operate in different regions of the state space with
different `local' transformation groups. To do this we introduce, as
in the previous sections, $C^{\infty }$ functions defined on
disjoint subsets of $\mathbb{R}^{n}$ as follows. Let $U_{i},\;1\leq
i\leq K$ (for some finite $K$) be bounded open subsets of
$\mathbb{R}^{n}$ such that there exist disjoint open neighbourhoods
$ V_{i}$ of $U_{i}$ for which
\[
V_{i}\supseteq \overline{{U}_{i}}\text{ and }V_{i}\cap
V_{j}=\emptyset ,\qquad i\neq j,\quad 1\leq i,j\leq K.
\]
Let $\varphi _{i}$ be a $C^{\infty }$ function such that
\[
\varphi _{i}(x)=\left\{
\begin{array}{ll}
1 & ,\;\text{if }x\in U_{i} \\
0 & ,\;\text{if }x\in \mathbb{R}^{n}\backslash V_{i}.
\end{array}
\right.
\]
Now let $G^{i},\;1\leq i\leq K$ be $K$ (linear) transformation
groups and let $t\rightarrow G^{i}(t)$ be $K$ smooth functions with
values in $G^{i}$. Then as in \textbf{Equation} \ref{first cinfinity
equation} we consider the system
\[
\dot{x}=f(x,t)
\]
and the transformed system
\begin{equation}
\dot{y}=\sum_{i=1}^{K}\varphi _{i}(y)\tilde{G}^{i}(y,t)
\end{equation}
where
\[
\tilde{G}^{i}(y,t)=\dot{G}^{i}(t)(G^{i})^{-1}(t)y+G^{i}(t)f\big((G^{i})^{-1}(t)y,t\big).
\]
Consider the effect of $G^{i}$ on $V_{i}$ at $t=\pi$. Define
\[
W_{i}=G^{i}(\pi)V_{i}
\]
and let
\[
X_{ij}=V_{i}\cap W_{j},\qquad 1\leq i,j\leq K
\]
be the $K^{2}$ intersections of the sets $\{V_{i}\}$ and
$\{W_{i}\}$. We assume that the functions $t\rightarrow G^{i}(t)$
are chosen so that the $X_{ij}$ are mutually disjoint. Let $\varphi
_{ij}$ be the obvious restriction of $\varphi _{i}$ to $X_{ij}$ and
consider the system
\begin{equation}
\dot{y}=\sum_{i=1}^{K}\varphi _{ij}(y)\tilde{G}^{i}(y,t).
\end{equation}
\begin{theorem}
Using the above notation, if the function
\[
\sum_{i=1}^{K}\varphi _{ij}(y)\tilde{G}^{i}(y,t)
\]
is periodic with period $\pi$, then the system
\begin{eqnarray*}
\dot{y} &=&\sum_{i=1}^{K}\varphi _{ij}(y)\tilde{G}^{i}(y,t),\qquad
y\in U,\;0\leq t\leq \pi \\
\dot{z} &=&2\sqrt{1-z^{2}},\qquad z(0)=0
\end{eqnarray*}
(where $U$ is a ball containing all sets $V_{i}$), is naturally
defined on the torus $U\times [0,1]/\sim $ where $\sim $ is as in
\textbf{Theorem} \ref{simple}.  \qquad  $\Box$
\end{theorem}

\begin{example}
\emph{We will use this method to generate systems with arbitrarily
knotted chaos. Consider first a system with unknotted chaos. Let
$U_{1},U_{2}$ be the sets}
\begin{eqnarray*}
U_{1} &=& \{(x_{1},x_{2}):0<x_{1}<1,0<x_{2}<\frac{1}{3}\} \\
U_{2} &=& \{(x_{1},x_{2}):0<x_{1}<1,\frac{2}{3}<x_{2}<1\}
\end{eqnarray*}
\emph{and $W_{1},W_{2}$ the sets}
\begin{eqnarray*}
W_{1} &=& \{(x_{1},x_{2}):0<x_{1}<\frac{1}{3},0<x_{2}<1\} \\
W_{2} &=& \{(x_{1},x_{2}):\frac{2}{3}<x_{1}<1,0<x_{2}<1\}.
\end{eqnarray*}
\emph{The transformation groups $G_{1},G_{2},G_{3}$ correspond to:\\
the `stretch and squeeze'}
\[
G_{1}(t):(x_{1},x_{2})\rightarrow (\frac{x_{1}}{t},tx_{2}),
\]
\emph{rotation}
\[
G_{2}(t):(x_{1},x_{2})\rightarrow
(t(x_{2}-2.5),-t(x_{1}-\frac{1}{6}))
\]
\emph{and translation}
\[
G_{3}(t):(x_{1},x_{2})\rightarrow (x_{1}+t,x_{2}-t).
\]
\emph{Finally we define}
\[
X_{ij}=V_{i}\cap W_{j},\qquad 1\leq i,j\leq 2,
\]
\emph{and we have the system}
\begin{eqnarray*}
\dot{y} &=& \sum_{i,j=1}^{2}\varphi _{ij}(y)\tilde{G}^{i}(y,t) \\
\dot{z} &=& 2\sqrt{1-z^{2}}
\end{eqnarray*}
where $\varphi _{ij}(y)$ is a $C^{\infty }$ function corresponding
to $X_{ij}$,
\[
G^{1}=G_{1},\qquad G^{2}=G_{3}\circ G_{2}\circ G_{1}
\]
\emph{and $\tilde{G}^{i}$ is obtained from $G^{i}$ as in
\textbf{Equation} \ref{second cinfinity equation}. This system has
chaotic orbits as shown in \textbf{Fig} \ref{smale horseshoe}. Note
that if $G_{i}(t),\quad 1\leq i\leq 3$ are properly chosen, the
system has no homoclinic orbits. (This simply implements the
\emph{`Smale horseshoe'} map.)}

\begin{figure}[!h]
\begin{center}
\includegraphics[width=4in]{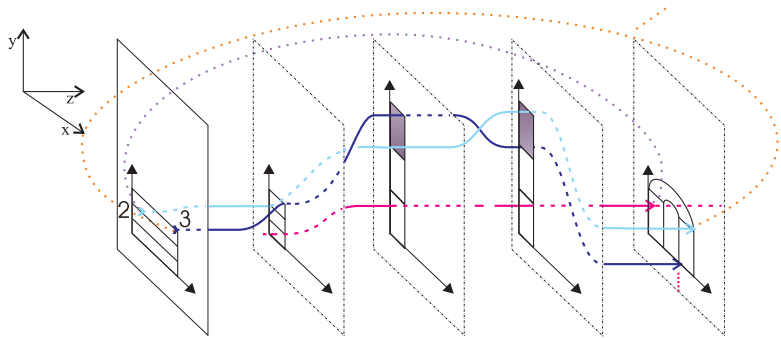}
\caption{Creating a chaotic system from the \emph{Smale horseshoe}}
\label{smale horseshoe}
\end{center}
\end{figure}

\emph{Now consider a system of the form}
\begin{eqnarray} \label{braid presentation for a system}
\dot{x} &=&f(x,t) \\
\dot{z} &=&2\sqrt{1-z^{2}}  \nonumber
\end{eqnarray}
\emph{defined for $x\in U$, where $U$ is some bounded open set in
$\mathbb{R}^{n}$. Let $\psi :[0,\pi ]\rightarrow \mathbb{R}^{n}$ be
any $C^{\infty }$ function
(which represents a strand of a braid) and let $\varphi :\mathbb{R}%
^{n}\rightarrow \Bbb{R}$ be a $C^{\infty }$ function which is 1 on
$U$ and 0 outside some neighbourhood of $U$. Then, if we put}
\[
y=x+\psi
\]
\emph{the system}
\begin{eqnarray*}
\dot{y} &=&\dot{x}+\dot{\psi}=(f(x,t)+\dot{\psi})\varphi (x-\psi (t)) \\
&=&\left( f(y-\psi ,t)+\dot{\psi}\right) \varphi (y)
\end{eqnarray*}
\emph{has trajectories like those of \textbf{Equation} \ref{braid
presentation for a system} in $U$, but `bent' by $\psi $. (See
\textbf{Fig} \ref{twisted braid})}

\begin{figure}[!hbp]
\begin{center}
\includegraphics[width=3.5in]{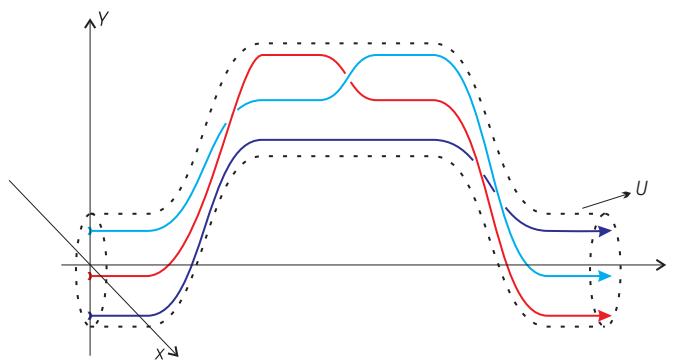}
\caption{A twisted braid} \label{twisted braid}
\end{center}
\end{figure}

\emph{More generally, if $U_{i},\;1\leq i\leq K$, are several open
(disjoint) sets in $\mathbb{R}^{n}$, and $\varphi _{i},\psi _{i}$
are associated functions as above, then the system}
\begin{eqnarray*}
\dot{y} &=&\sum_{i}\left( f(y-\psi _{i}(t),t)+\dot{\psi}_{i}(t)\right)
\varphi _{i}(x-\psi _{i}(t)) \\
\dot{z} &=&2\sqrt{1-z^{2}}
\end{eqnarray*}
\emph{will have trajectories similar to a given system in the
regions $U_{i},$but `bent' by the functions $\psi _{i}$. Clearly, by
appropriate choice of $\psi _{i}$ and $\varphi _{i}$ we can obtain a
system with arbitrarily knotted chaos, which contains no homoclinic
orbits. \textbf{Fig} \ref{knotted chaos} shows a braid
representation of a \emph{trefoil} knot which contains a chaotic
system, \emph{Smale horseshoe}, inside.}

\begin{figure}[!hbp]
\begin{center}
\includegraphics[width=4in]{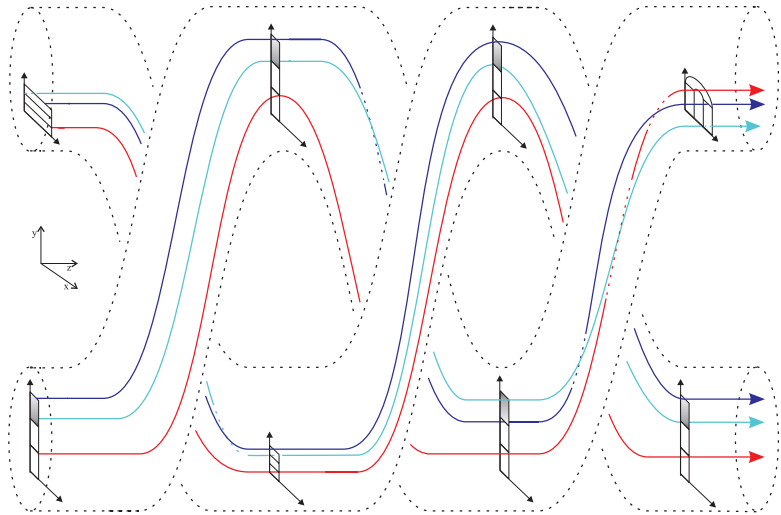}
\caption{Knotted chaos} \label{knotted chaos}
\end{center}
\end{figure}

\end{example}

\section{Conclusions}

In this paper we have shown how to generate three-dimensional
systems containing arbitrarily knotted chaos by using the theory of
transformation groups and $C^{\infty }$ functions. By `twisting' a
simple existing dynamical system by local transformation groups and
making the resulting system periodic, virtually any dynamical
behaviour can be obtained. In the second part of the paper we shall
consider more general three-manifolds and dynamical systems defined
on them by using the theory of \emph{Heegaard splittings}. Every
three-manifold has a \emph{Heegaard splitting} which represents it
in the form of two three manifolds with genus $p$ surfaces glued
together along a framed knot.

\end{document}